\documentclass{iau} 
\usepackage{graphicx}
\usepackage{natbib}

\newcommand\mnras{\textit{MNRAS}}
\newcommand\apj{\textit{ApJ}}

\newcommand\apjs{\textit{ApJ}}
\newcommand\aj{\textit{AJ}}

\newcommand\procspie{\textit{Proc. of the SPIE}}

\title[Stellar multiplicity in the Milky Way Galaxy] 
{Stellar multiplicity in the Milky Way Galaxy}

\author[E. Stonkut\.{e} et al.]   
{E. Stonkut\.{e}$^{1,2}$, R. P. Church$^2$, S. Feltzing$^2$
\and J.~A.~Johnson$^3$}
\affiliation{$^1$Institute of Theoretical Physics and Astronomy, Vilnius University, Saul\.{e}tekio al. 3, LT-10222, Vilnius, Lithuania. Email: {\tt edita.stonkute@tfai.vu.lt} \\[\affilskip]
$^2$Lund Observatory, Department of Astronomy and Theoretical Physics, Box 43, SE-22100, Lund, Sweden\\
$^{3}$The Ohio State University, Columbus, OH 43210, USA\\}

\pubyear{2017}
\volume{334}  
\jname{Rediscovering our Galaxy}
\editors{C. Chiappini, I. Minchev, E. Starkenburg, \& M. Valentini, eds.}
\begin{document}

\maketitle

\begin{abstract}
We present our models of the effect of binaries on high-resolution spectroscopic surveys. We want to determine how many binary 
stars will be observed, whether unresolved binaries will contaminate measurements of chemical 
abundances, and how we can use spectroscopic surveys to better constrain the population of binary stars in the Galaxy.  
Using a rapid binary-evolution algorithm that enables modelling of the most complex binary systems we generate a series of 
large binary populations in the Galactic disc and evaluate the results. As a first application we use our model to study the binary fraction in APOGEE giants.
We find tentative evidence for a change in binary fraction with metallicity. 
\end{abstract}

\firstsection 
\section{Introduction}

Stellar multiplicity leads to many interesting astrophysical phenomena, for
example, gravitational wave sources, gamma-ray bursts,  and type Ia
supernovae.
It is known that 34\% $ \pm$ 2\% of solar-type (F6--K3) stars in the Solar Neighbourhood  are in binaries \citep{Raghavan10}, but the
 frequency of binary stars  outside the Solar Neighbourhood is uncertain.
For ongoing and up-coming large 
spectroscopic surveys, such as RAVE \citep{Kunder17}, APOGEE \citep{Majewski15}, {\sl{Gaia}}-ESO \citep{Gilmore12}, 
GALAH \citep{DeSilva15}, LAMOST \citep{Deng12} or 4MOST \citep{deJong16} it is important to identify as well as quantify the binaries to clean the survey 
products from potentially faulty results.

\section{The effect of binaries on high-resolution spectroscopic surveys}

\begin{figure}

\begin{center}
 \includegraphics[width=5.2in]{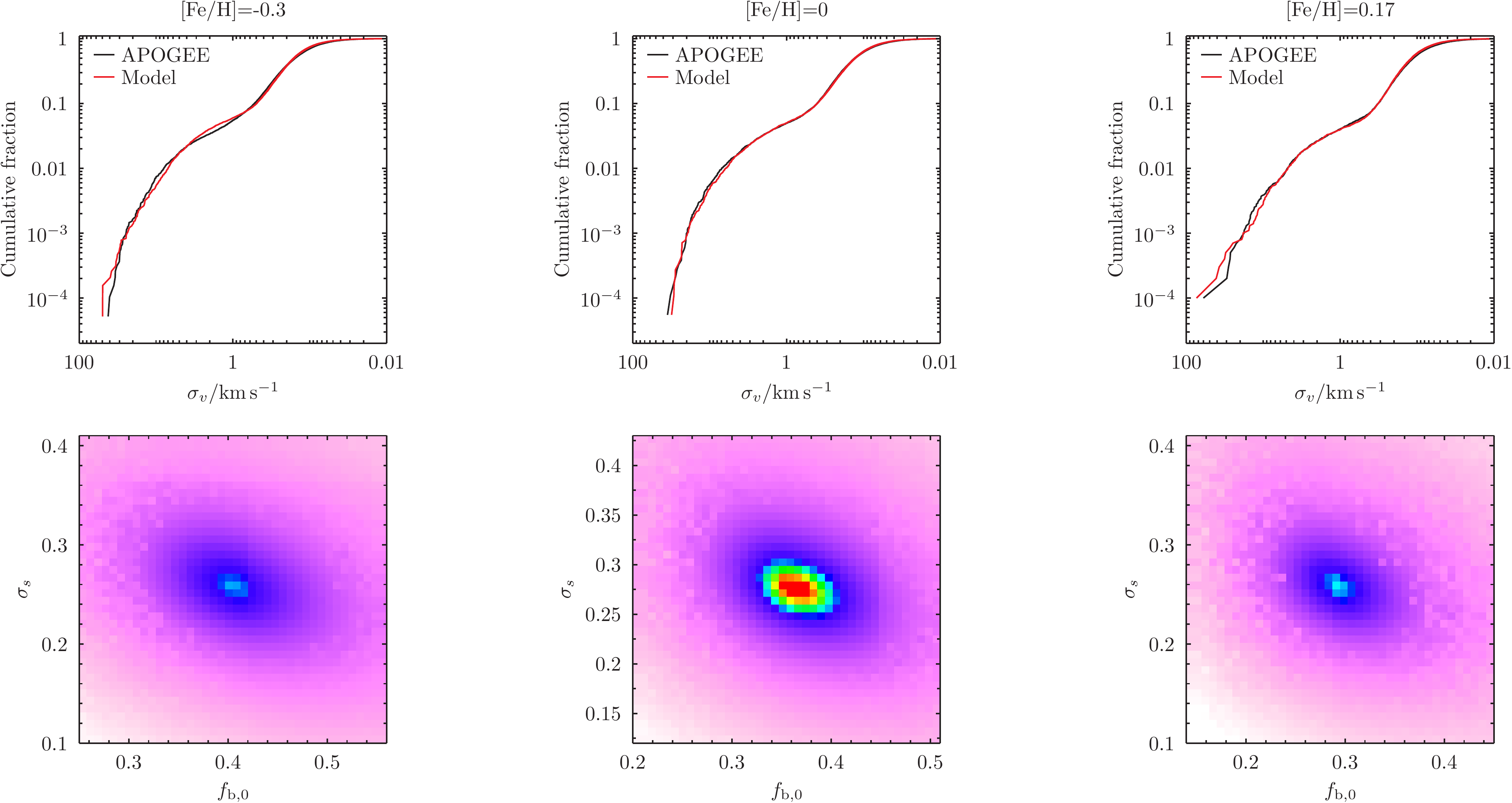} 
 \caption{Top: the best fit cumulative distributions of the velocity dispersion of the simulated binaries (red line) and the observed velocity dispersion from APOGEE DR\,12  (black line) in three metallicity bins \rm{[Fe/H]}=$(-0.3;0.0; +0.17$). Bottom: initial binary fraction $f_{b,0}$ versus the nuisance parameter $\sigma_{s}$, that describes the intrinsic scatter of the APOGEE measurements.}
   \label{fig1}
\end{center}
\end{figure}

As an application we make mock APOGEE observations of red giants and subgiants. Our stars selection function mimics the selection function of APOGEE.
We selected Galactic disc stars from APOGEE DR\,12 where Galactic longitude: 24$^{\circ}$ $\leq$ $l $ $\leq$ 240$^{\circ}$ and latitude: $|b|$ $\leq$ 16$^{\circ}$;
the signal-to-noise ratio of individual spectra\,$\geq$\,20; the effective temperature: 3500\,$\leq$\,Teff\,$\leq$\,5500, [K] and \rm{log(g)}\,$\leq$\,4.0, [cgs]; and stars that have been visited more than two times: Nvists\,$>$\,2.

Binary and single star evolution is performed by the rapid binary-star evolution ({\sc{bse}}) algorithm \citep{Hurley02}.
We assume that the initially more massive stars in the binary have masses between $0.9\,{\rm M}_{\odot}\le m_1 \le 100\,{\rm M}_{\odot}$. 
We generate $m_1$ from the initial mass function of \citet{Kroupa93}.
The mass of the companion is between $0.1\,{\rm M}_{\odot} \le m_{2} \le 100$ and is drawn assuming the mass ratio distribution is flat in $q$, where $q=m_{2}/m_{1}$\,$\le$\,1.
Our next assumption is that our binary stars in the Galactic disc are in three metallicity bins \rm{[Fe/H]}=$(-0.3;0.0; +0.17$) and have broad uniform age distribution from 0 to 10\,Gyr.
The distribution of binary periods, $P$, is log-normal with $\overline{\log P}$ = 4.8 and $\sigma$$_{\log P}$ = 2.3, here the orbital period is in days.
The distribution of the orbital eccentricity ($e$)  is chosen to be dynamically relaxed (thermal)$f(e)\varpropto2e$.

\section{Results}

In Fig. 1 
the cumulative functions suggest that the model fits the observations well, which is
consistent with most of the stars with high velocity scatter being binaries, and with the binary population being the same elsewhere in the Galaxy.
The estimated initial binary fraction ($f_{b,0}$) for \rm{[Fe/H]}\,=\,0.0 is 36\% which is consistent with  \citet{Raghavan10} for solar-type stars.
There is evidence for $f_{b,0}$ decreasing with increasing \rm{[Fe/H]} consistent with other studies \citep[e.g.][]{Yuan15}.

We intend to investigate to what extent we can constrain the frequency of binaries, and whether we can detect any systematic variation with metallicity in the Galaxy. 
The detailed results will be presented in Stonkut\.{e}, Church \& Feltzing 2017 (in prep.).

\

{\it{Acknowledgments}}.
E.S., R.C. and S.F were supported by the project grant ``The New Milky Way'' from Knut and Alice Wallenberg Foundation.

\end{document}